\documentstyle[11pt,newpasp,twoside,graphicx,natbib]{article}
\bibpunct[]{(}{)}{;}{a}{}{,}

\def\edcomment#1{\iffalse\marginpar{\raggedright\sl#1\/}\else\relax\fi}
\reversemarginpar

\begin{document}
\title{H$_2$-bearing damped Lyman-$\alpha$ systems as tracers of
  cosmological chemical evolution}

\author{M.~T.~Murphy,}
\affil{Institute of Astronomy, Madingley Road, Cambridge CB3 0HA, UK}
\author{S.~J.~Curran, J.~K.~Webb}
\affil{School of Physics, University of New South Wales, Sydney 2052,
  Australia}

\begin{abstract}\begin{sloppypar}
The chemical abundances in damped Lyman-$\alpha$ systems (DLAs) show more
than 2 orders of magnitude variation at a given epoch, possibly because
DLAs arise in a wide variety of host galaxies. This could significantly
bias estimates of chemical evolution. We explore the possibility that DLAs
in which H$_2$ absorption is detected may trace cosmological chemical
evolution more reliably since they may comprise a narrower set of physical
conditions. The 9 known H$_2$ absorption systems support this hypothesis:
metallicity exhibits a faster, more well-defined evolution with redshift
than in the general DLA population. The dust-depletion factor and,
particularly, H$_2$ molecular fraction also show rapid increases with
decreasing redshift. We comment on possible observational selection effects
which may bias this evolution. Larger samples of H$_2$-bearing DLAs are
clearly required and may constrain evolution of the UV background and DLA
galaxy host type with redshift.
\end{sloppypar}\end{abstract}

\section{Introduction}\label{s:intro}

Apart from providing independent supporting evidence of the big bang, the
detection and subsequent study of cosmological chemical evolution provides
the empirical details of galaxy formation and evolution. How primordial and
processed gas is consumed by star formation, the dominant feedback
processes and merging scenarios, may all contribute to the overall
evolution of chemical abundances. One high-precision probe of this
evolution is the spectroscopic study of damped Lyman-$\alpha$ systems
(DLAs): absorbers with neutral hydrogen column densities $N({\rm H{\sc
\,i}}) \geq 2\times 10^{20}{\rm \,cm}^{-1}$. Although these observations
demonstrate that DLAs arise along lines of sight through distant galaxies,
they do not directly disclose details such as the galaxy's morphology,
luminosity, mass or age.

There exists substantial evidence that DLAs arise in a variety of galaxy
types. At low-$z$ ($z_{\rm abs} \la 1.5$), kinematic and H{\sc \,i} 21-cm
absorption studies (e.g.~\citealt{BriggsF_85a}; \citealt*{BriggsF_01a})
suggest a significant contribution from spiral galaxies, a view supported
at high-$z$ by kinematic modelling and abundance studies
\citep{ProchaskaJ_97b,WolfeA_99a}. However, direct imaging at low-$z$
\citep[e.g.][]{LeBrunV_97a} reveals that DLA hosts are a mix of irregulars,
spirals and low surface-brightness galaxies (LSBs). A recent $z=0$ H{\sc
\,i} 21-cm emission study \citep*{Ryan-WeberE_03a} supports
this. \citet{BoissierS_03b} argue that the number of DLAs per redshift
interval and the $N({\rm H{\sc \,i}})$ distribution imply that DLAs at
$z<2$ are a mix of spirals and LSBs whereas, at higher $z$, they are more
likely to be dwarfs. This is supported by fitting of chemical evolution
models to DLA metal abundances \citep[e.g.][]{BakerA_00a} and by recent
21-cm absorption searches at high redshift \citep{KanekarN_03a}. Though
further work is clearly needed, the direct and indirect evidence for a
`mixed bag' of DLA hosts is already compelling.

\section{Selecting a less biased tracer of chemical evolution}\label{s:select}

\begin{figure}[ht!]
\begin{center}
\includegraphics[width=0.70\textwidth]{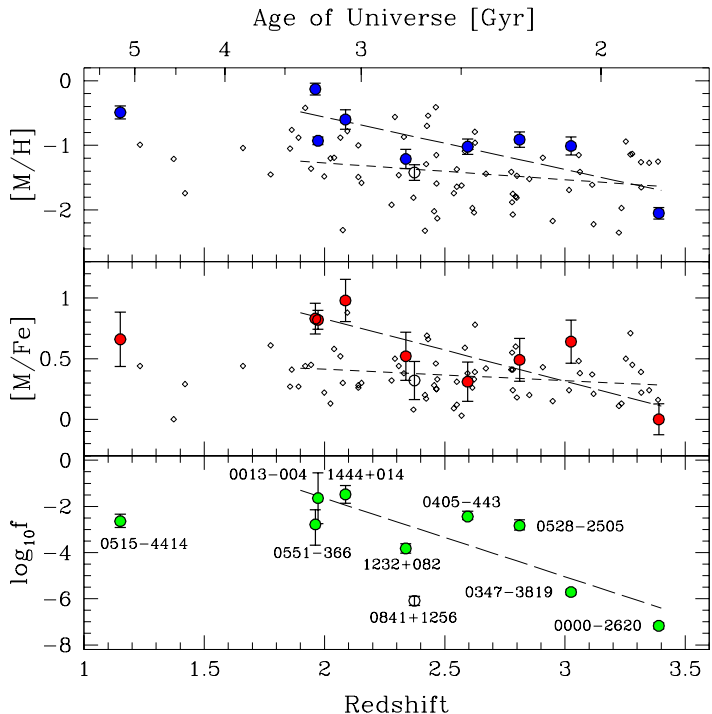}
\end{center}
\vspace{-0.3cm}
\caption{Metallicity [M/H], dust-depletion [M/Fe], and molecular fraction
  ($f$) evolution for confirmed (filled circles) and tentative (hollow
  circle) H$_2$ detections compared with values from the
  \citet{ProchaskaJ_03b} DLAs (small diamonds). The values adopted are
  discussed in the text. Linear fits to the H$_2$-bearing DLAs (long
  dashes) and general DLAs (short dashes) were obtained by quadrature
  addition of a constant to the individual error bars such that
  $\chi^2_\nu=1$. Due to its much lower $z_{\rm abs}$ and potentially high
  photo-dissociation rate \citep{ReimersD_03a}, we do not include
  0515$-$4414 in these fits (but see Table \ref{tab:1}).}
\label{fig:1}
\end{figure}

Evidence for an increase in DLA metallicities, [M/H], with cosmic time has
emerged only gradually
\citep{PettiniM_95a,LuL_96b,VladiloG_00a,KulkarniV_02a,ProchaskaJ_03b}, the
latter reference using over 100 DLAs to provide the strongest statistical
evidence so far. Such large samples are required due to the huge scatter
($\sim\!2{\rm \,dex}$) in [M/H] at a given epoch (see Fig.~\ref{fig:1}), a
diversity expected given the variety of DLA hosts discussed above. However,
the diversity in [M/H] could also significantly {\it bias} any estimate of
chemical evolution, as could several observational selection effects
\citep[e.g.][]{HouJ_01a}.

In \citet{CurranS_04c} we suggest that by selecting those DLAs in which
H$_2$ absorption is detected, one may reduce or possibly avoid some of the
biases besetting the general DLA population. There are currently 9
confirmed H$_2$-bearing DLAs (see below) and, typically, H$_2$ is detected
in only a few velocity components. These H$_2$-bearing components seem
distinct from the others, showing lower temperatures and higher dust
depletion factors [M/Fe] \citep{PetitjeanP_02a}. Therefore, H$_2$-bearing
DLAs might be a less biased tracer of chemical evolution than the general
DLA population since they may allow one to focus on a narrower range of
physical conditions throughout cosmic time.

\subsection{Known H$_2$-bearing DLAs}\label{ss:data}

H$_2$ is detected in DLAs via the Lyman and Werner-band UV absorption lines
which generally lie in the H{\sc \,i} Lyman-$\alpha$ forest. A compilation
of results from H$_2$ searches in DLAs is given in table 8 of \citet*[][,
hereafter L03]{LedouxC_03a}, for which 7 DLAs have confirmed H$_2$
detections and metallicity measurements: 0013$-$004 ($z_{\rm abs}=1.973$),
0347$-$3819, 0405$-$443 ($z_{\rm abs}=2.595$), 0528$-$2505, 0551$-$366,
1232$+$082 and 1444$+$014. The DLA towards 0013$-$004
\citep*{PetitjeanP_02a} comprises several absorbing components, of which 2
are dominant (their components $c$ and $d$): $N({\rm H{\sc \,i}})$ is
measured at the mean redshift and we use the mean $N($H$_2)$ with error
given by the range of $N($H$_2)$ in L03. For the DLA towards 1232$+$082 we
use the $N($H$_2)$ value and error from \citet*{SrianandR_00a}.

We include two further DLAs: (i) The recent detection towards 0515$-$4414
\citep{ReimersD_03a} and (ii) 0000$-$2620, regarded as only a tentative
detection by L03. However, the H$_2$ identification has been carefully
scrutinised by \citet{LevshakovS_00c,LevshakovS_01b} and relies on two
H$_2$ absorption features, the L(4-0)R1 and W(2-0)Q(1) lines, the former
appearing relatively clean from Lyman-$\alpha$ forest blending. A tentative
detection of H$_2$ in 0841$+$1256 has also been reported
\citep{PetitjeanP_00a}, though confirmation requires future data and
analyses.

\subsection{Results}\label{ss:resu}

\begin{table}
\caption{Statistics for correlations in Fig.~\ref{fig:1}. $P(\tau)$ is the
  probability of a chance correlation using a Kendall's-$\tau$ test, $B$ is
  the slope (with 1\,$\sigma$ error) of the best fit, $F$ is the F-test
  statistic (ratio of variances of two distributions) derived after
  subtracting the best fits from the data, and $P(F)$ is the probability
  that $F$ could be exceeded by chance alone. Since 0515$-$4414 is at much
  lower $z_{\rm abs}$ and may have a much higher photo-dissociation rate
  than the other H$_2$-bearing DLAs \citep{ReimersD_03a}, we present
  separate statistics for samples excluding and including 0515$-$4414. Note
  that this makes little difference.}
\smallskip
\begin{center}
\begin{tabular}{lcccccc}\tableline
                    & \multicolumn{2}{c}{H$_2$-selected DLAs} & \multicolumn{2}{c}{`General' DLAs} & \multicolumn{2}{c}{F-test} \\
Sample              & $P(\tau)$ & $B$                         & $P(\tau)$ & $B$                    & $F$  & $P(F)$              \\\tableline
\multicolumn{7}{l}{{\bf [M/H] vs.}~{\it\bf z}$_{\rm\bf abs}$} \\
Exc.~0515           & 0.08      & $-0.81\!\pm\!0.27$          & 0.14      & $-0.26\!\pm\!0.15$     & 2.2  & 0.28                \\
Inc.~0515           & 0.04      & $-0.60\!\pm\!0.20$          & 0.03      & $-0.26\!\pm\!0.11$     & 1.9  & \,\,0.33            \vspace{0.05cm}\\
\multicolumn{7}{l}{{\bf [M/Fe] vs.}~{\it\bf z}$_{\rm\bf abs}$} \\
Exc.~0515           & 0.05      & $-0.51\!\pm\!0.12$          & 0.24      & $-0.09\!\pm\!0.06$     & 1.7  & 0.47                \\
Inc.~0515           & 0.06      & $-0.36\!\pm\!0.13$          & 0.25      & $-0.07\!\pm\!0.04$     & 1.0  & \,\,0.95            \vspace{0.05cm}\\
\multicolumn{7}{l}{{\it\bf f} {\bf vs.}~{\it\bf z}$_{\rm abs}$} \\
Exc.~0515           & 0.04      & $-3.40\!\pm\!0.86$          & ---       & ---                    & ---  & ---                 \\
Inc.~0515           & 0.06      & $-2.00\!\pm\!0.75$          & ---       & ---                    & ---  & ---                 \vspace{0.05cm}\\
\tableline\tableline
\end{tabular}
\end{center}
\label{tab:1}
\vspace{-0.7cm}
\end{table}

In Fig.~\ref{fig:1} we plot against $z_{\rm abs}$ the metallicity [M/H],
dust-depletion factor [M/Fe] and molecular fraction $f\equiv 2N({\rm
H}_2)/[N({\rm H{\sc \,i}}) + 2N({\rm H}_2)]$ for the 9 H$_2$-bearing DLAs,
comparing the former two quantities with DLAs in \citet{ProchaskaJ_03b}
with Zn, Si, S or O metallicity over the relevant redshift range. [M/H] for
the H$_2$-bearing DLAs is [Zn/H], except for 0347$-$3819 and 1232$+$082
where it is [S/H] and [Si/H]. For 0515$-$4414, [Zn/Fe] is from
\citet{delaVargaA_00a}.

The main results from Fig.~\ref{fig:1} are summarized by the statistics in
Table \ref{tab:1}: (i) [M/H], [M/Fe] and $f$ for the H$_2$-bearing DLAs are
anti-correlated with $z_{\rm abs}$ at the 95\% confidence level (i.e.~more
significant than for the general DLA population), (ii) [M/H] shows a
steeper evolution with $z_{\rm abs}$ and a smaller scatter about the slope
than the general DLA population, (iii) [M/Fe] in H$_2$-bearing DLAs shows
strong evolution with $z_{\rm abs}$ while the general DLAs show no evidence
for evolution, and (iv) $f$ ranges over $6{\rm \,dex}$ and shows a very
steep evolution with $z_{\rm abs}$. The new results (i)--(iii) support our
hypothesis that H$_2$-bearing DLAs form a chemically distinct sub-class and
may trace chemical evolution more reliably. The $f$--$z_{\rm abs}$
correlation was studied by L03 and is discussed further below.

The [M/H]s and $f$s in Fig.~\ref{fig:1} are measured using the total
$N({\rm H{\sc \,i}})$ across the DLA profile and are not specific to the
H$_2$-bearing components. [M/Fe] is generally found to be uniform across
most DLA profiles \citep{ProchaskaJ_03a}, indicating that [M/H] should be
uniform. However, the H$_2$-bearing components typically have much higher
[M/Fe] values than other components in the same DLA (e.g.~L03 and
0347$-$383's [Si/Fe] profile in \citealt{ProchaskaJ_03a}). These components
usually dominate the non-refractory metal-line profiles and so, although
[M/H] and $f$ will be systematically underestimated, the effect will not be
large. The fitted slopes in Fig.~\ref{fig:1} are likely to be reasonably
robust against this effect, but a larger sample and more detailed study is
clearly required.

What observational selection effects and biases could contribute to the
steep $f$--$z_{\rm abs}$ evolution? Firstly, the sample is inhomogeneous
since the spectra do not all have similar S/N and since the H$_2$ detection
methods and criteria were not uniform. Indeed, the weak H$_2$ lines
detected towards 0000$-$2620 are at the typical non-detection level (see
fig.~16 in L03). The $f$--$z_{\rm abs}$ correlation is therefore only
tentative. Secondly, the H$_2$ detection limit will alter with $z$:
equivalent widths increase but Lyman-$\alpha$ forest blending worsens with
increasing $z$. Though this is an unlikely culprit for the $6{\rm \,dex}$
evolution observed, precise quantification of these competing effects
requires numerical simulations. DLAs containing large amounts of dust could
suppress detection of their background quasars and may therefore be
`missing' from our sample \citep*{FallS_89a}. However, since [M/H], [M/Fe]
and $f$ are positively correlated with each other (e.g.~L03), this effect
is likely to suppress, rather than create, the correlations in
Fig.~\ref{fig:1}. A recent survey for DLAs towards radio-selected quasars
\citep{EllisonS_01c} also indicates that the number of such `missing'
quasars is likely to be small.

\vspace{-0.1cm}
\section{Interpretations and conclusions}\label{s:conc}

Selecting those DLAs which exhibit H$_2$ absorption may focus on systems
with a narrower range of physical conditions than the DLA population as a
whole. Tentative support for this conjecture lies in the steeper, tighter
[M/H]--$z_{\rm abs}$ anti-correlation observed for the H$_2$ systems
studied here. \citet*{HouJ_01a} recently presented detailed chemical
evolution models which give a slope for the [M/H]--$z_{\rm abs}$ relation
of $\sim\!-0.6{\rm \,dex}$. They correct this result for various
observational biases to match the shallower slope observed for the general
DLA population. However, the steep [M/H]--$z_{\rm abs}$ slope observed for
H$_2$-bearing systems could be less affected by these biases and may avoid
those introduced by sampling many different ISM gas phases. It might
therefore be more comparable to the uncorrected slopes in the
models. H$_2$-selected DLAs are therefore a candidate for a less biased
probe of chemical evolution.

That there exists such a large range ($\sim\!6{\rm \,dex}$) in the values
of $f$ in Fig.~\ref{fig:1} may not be surprising: \citet{SchayeJ_01b}
describes a ${\rm [Zn/H]}=-1$ photo-ionization model for clouds in local
hydrostatic equilibrium. For a representative incident UV background flux
and dust-to-metals ratio, the molecular fraction in this model shows a
sudden increase of $\sim\!4{\rm \,dex}$ for only a small increase in the
total hydrogen density. Therefore, the very steep $f$--$z_{\rm abs}$
correlation could be achieved with a modest increase in dust content at
lower redshifts, consistent with the observed [M/Fe]--$z_{\rm abs}$
anti-correlation. L03 also discuss how $f$ might be very sensitive to local
physical conditions. For example, within the Schaye model, one expects an
anti-correlation between $f$ and the intensity of the UV
background. However, the behaviour of the UV background flux with redshift
over the range $z=\!1$--3 is still a matter of considerable
uncertainty. The strong decrease in $f$ at high redshift may also be
consistent with recent H{\sc \,i} 21-cm absorption measurements in DLAs
\citep{KanekarN_03a}, where a generally higher spin/excitation temperature
is found at $z>2.5$. With an increased sample size and more detailed
analyses, the $f$--$z_{\rm abs}$ anti-correlation, if real, may provide
complementary constraints on these problems.

\acknowledgments{We thank Ed Jenkins, Charley Lineweaver and Mark Whittle
for discussions. MTM thanks the IAU for a generous travel grant and PPARC
for support at the IoA.}


\end{document}